\documentclass[useAMS,usenatbib]{mn2e}
\usepackage{graphicx,amssymb,color}
\usepackage[normalem]{ulem}

\newcommand{\rev}{ }

\title[History of WD~J0914+1914~b]
{The dynamical history of the evaporating or disrupted ice giant planet around white dwarf WD~J0914+1914}

\author[]{Dimitri Veras$^{1,2}$\thanks{E-mail: d.veras@warwick.ac.uk}\thanks{STFC Ernest Rutherford Fellow},
Jim Fuller$^3$
\\
$^{1}$Centre for Exoplanets and Habitability, University of Warwick, Coventry CV4 7AL, UK
\\
$^{2}$Department of Physics, University of Warwick, Coventry CV4 7AL, UK
\\
$^{3}$TAPIR, Mailcode 350-17, California Institute of Technology, Pasadena, CA 91125, USA
}

\pubyear{2020}

\begin{document}
\label{firstpage}
\pagerange{\pageref{firstpage}--\pageref{lastpage}}
\maketitle

\begin{abstract}
Robust evidence of an ice giant planet shedding its atmosphere around the white dwarf WD~J0914+1914 represents a milestone in exoplanetary science, allowing us to finally supplement our knowledge of white dwarf metal pollution, debris discs and minor planets with the presence of a major planet. Here, we discuss the possible dynamical origins of this planet, WD~J0914+1914~b. The very young cooling age of the host white dwarf (13 Myr) combined with the currently estimated planet-star separation of about 0.07 au imposes particularly intriguing and restrictive coupled constraints on its current orbit and its tidal dissipation characteristics. The planet must have been scattered from a distance of at least a few au to its current location, requiring the current or former presence of at least one more major planet in the system {\rev in the absence of a hidden binary companion}. We show that WD~J0914+1914~b could not have subsequently shrunk its orbit through chaotic f-mode tidal excitation (characteristic of such highly eccentric orbits) unless the planet was or is highly inflated and {\rev possibly} had partially thermally self-disrupted from mode-based energy release. We also demonstrate that if the planet is currently assumed to reside on a near-circular orbit at 0.07 au, then non-chaotic equilibrium tides impose unrealistic values for the planet's tidal quality factor. We conclude that WD~J0914+1914~b either {\rev (i) actually resides interior to 0.07 au, (ii) resembles a disrupted ``Super-Puff'' whose remains reside on a circular orbit, or (iii) resembles a larger or denser ice giant on a currently eccentric orbit}.  Distinguishing these {\rev three} possibilities strongly motivates follow-up observations.
\end{abstract}


\section{Introduction}

The first exoplanetary system signatures discovered around main-sequence stars arose from
major planets \citep{cametal1988,latetal1989,hatcoc1993,mayque1995, marbut1996}. In a twist of fate, the situation 
for white dwarf planetary systems is just the opposite: secure evidence for a major
planet \citep{ganetal2019} represents one of the {\it last} significant missing components of these systems to be found.

Over the past century \citep{vanmaanen1917,vanmaanen1919}, mounting discoveries of
planetary remnants in white dwarf atmospheres have led to surveys which show that 25-50 per
cent of all single Milky Way white dwarfs are metal polluted with planetary materials \citep{zucetal2003,zucetal2010,koeetal2014}. This debris is predominantly chemically consistent with fragments from rocky bodies 
{\rev \citep{zucetal2007,ganetal2012,juryou2014,haretal2018,holetal2018,doyetal2019,swaetal2019,xuetal2019,bonetal2020}}. Major rocky planets, however, are simply not numerous enough to represent the progenitors of these pollutants because they would not approach 
white dwarfs at a sufficiently high frequency \citep{veras2016a,veretal2016}. 

The debris needs to arise from larger reservoirs of minor planets, such as moons \citep{payetal2016,payetal2017} or
analogues to the asteroid belt \citep{debetal2012,frehan2014,smaetal2018,smaetal2019,antver2016,antver2019} or to the Kuiper belt \citep{bonetal2011,musetal2018,griver2019,makver2019};
comets are too volatile-rich and sparse to provide as good of a match with the debris \citep{alcetal1986,veretal2014a,stoetal2015,caihey2017}. However, without a stellar companion nor major planets, rocky minor planets cannot self-propel themselves into the 
white dwarf \citep{veretal2015a,veretal2018a}. Despite this need for major planets at distances within tens or hundreds of au, 
they had remained undetected\footnote{A near-exception includes an object residing at a distance of about 2500 au that was found orbiting WD 0806-661b and was classified as a possible brown dwarf by the authors \citep{luhetal2011}, despite harbouring a mass of approximately 10 Jupiter masses. Also, a different, circumbinary object of lower mass, PSR B1620-26b, was discovered orbiting both a pulsar and a white dwarf \citep{sigetal2003}.}. 

Concurrent with the increasing detections of white dwarf metal pollution were the first detection of a white
dwarf debris disc \citep{zucbec1987} and the first detection of an orbiting minor planet \citep{vanetal2015},
but still no major planet. Now over 40 discs are known \citep[e.g.][]{gaeetal2006,farihi2016,denetal2018} and a 
second \citep{manetal2019} and likely third minor planet {\rev \citep{vanetal2019,veretal2020}} have been detected orbiting different white dwarfs. Nevertheless, the formation of these discs \citep{debetal2012,veretal2014b,veretal2015b,malper2020a,malper2020b} still requires a major planet to perturb a minor planet into the white dwarf {\rev tidal disruption} radius \citep{veras2016a}. Also, the origin of the near-circular 
orbit {\rev \citep{guretal2017,veretal2017a,duvetal2020}} of the first minor planet found orbiting a white dwarf \citep{vanetal2015} remains one of the foremost unexplained problems in white dwarf planetary 
science, but envisaging an inward migration scenario without the help of a major planet is challenging.

A breakthrough arrived when \cite{ganetal2019} detected robust chemical signatures of 
accretion from evaporation of one of these major planets around the
metal-polluted and disc-bearing single white dwarf WD~J0914+1914. This detection was 
corroborated by several factors:

\begin{itemize}

\item The chemical absence of Earth-like compositions within the atmosphere of  WD~J0914+1914
is highly unusual for metal-polluted white 
dwarfs {\rev \citep{xuetal2017,haretal2018,holetal2018,doyetal2019,swaetal2019,xuetal2019,bonetal2020}}
and indicates a lack of rocky body accretion.

\item A pristine giant planet might be expected to contain methane in the outer layers of its atmosphere.
The claim that these outer layers have been evaporated is supported by the non-detection of carbon in 
WD~J0914+1914. Further, the upper limit implies a sub-solar carbon abundance.

\item Orbiting WD~J0914+1914 is a gas-only disc; the first to be discovered around a metal-polluted white dwarf.

\item The WD~J0914+1914 planetary system represents the first where independent measurements of 
absorption lines from the atmospheric metal pollution and emission lines from the metals in the gaseous 
disc were both obtained and match.

\end{itemize}

As a result, we know the composition of the planet's atmosphere better than its orbit; for main-sequence planets,
the opposite is nearly always true.
\cite{ganetal2019} estimated that the current location of the planet, although still very uncertain, is 
at a distance of about $15R_{\odot}\approx~0.070$~au. They further argued that the planet should be
further away than the outer extent of the disc, which is constrained through emission lines to be located at about 
$10R_{\odot}\approx~0.046$~au. However, in principle, a sufficiently low-mass
planet may be embedded within the disc and not have opened up a gap. Throughout this paper, we 
will keep in mind these uncertainties, but adopt $0.070$~au as the current planet distance for our computations.

Although the host star has a mass of $0.56~\pm~0.03 M_{\odot}$ (a common value; \citealt*{treetal2016}), its cooling
age of $13.3~\pm~0.5$~Myr  is very young compared to most known metal-polluted white dwarfs.
The term ``cooling age'' simply refers to the age of the star after it became a white dwarf. For WD~J0914+1914,
its cooling age represents the crucial dynamical constraint. A cooling age of $13.3$ Myr is considered so 
young because white dwarf planetary remnants have been observed in the atmospheres of white dwarfs with cooling
ages of 8 Gyr \citep{holetal2018}. \cite{verful2019} highlighted how such young cooling ages can place
constraints on the orbital history and tidal dissipation of gaseous planets. 

Here we apply their results
to the WD~J0914+1914 system, and henceforth use the designation WD~J0914+1914~b for the planet.
We address the orbital evolution of WD~J0914+1914~b in Section 2, briefly discuss our results in Section 3,
and conclude in Section 4.

\section{Orbital evolution}

\subsection{The planet's initial position}

WD~J0914+1914~b needed to survive the main sequence and giant branch stages of the
star's evolution before being relocated to a distance of about $0.07$ au. The planet's prospects
for survival during these phases relies on a combination of its initial position (where it formed) and details of
its host star evolution.

\cite{ganetal2019} claimed
that based on the current white dwarf mass of $0.56 \pm 0.03 M_{\odot}$, the star's progenitor 
main-sequence mass was $1.0-1.6M_{\odot}$. By considering the extremes of this range, 
we note that the evolution of $1.0M_{\odot}$ and $1.6M_{\odot}$ stars are 
qualitatively and quantitatively different.
A $1.0M_{\odot}$ star features comparably significant and potentially destructive (to planets) red giant
branch and asymptotic giant branch phases. Recent studies of the solar system's post-main-sequence evolution 
\citep{schsmi2008,veras2016b} revealed that the Sun's envelope will extend out to about 1 au at the tip
of both phases. In contrast, for $1.6M_{\odot}$ stars, the red giant envelope extends out to about 0.8 au
\citep{viletal2014} whereas the asymptotic giant envelope extends out to about 1.5 au \citep[Fig. 3 of][]{veras2016a}.

Numerous studies have now shown that the tidal interaction between giant branch stars and giant planets
require the latter to reside at distances of at least 1 au beyond the extent of the stellar envelope in order
to survive 
\citep{villiv2009,kunetal2011,musvil2012,adablo2013,viletal2014,madetal2016,staetal2016,galetal2017,raoetal2018,sunetal2018}.
Therefore, regardless of the progenitor mass or the stellar model used, a reasonable assumption 
is that WD~J0914+1914~b 
{\rev began} the white dwarf phase at a distance of {\it at least} 2-3 au.
Scenarios where the planet would instead acquire an orbital distance of just 0.07 au 
during a common envelope phase when the star ascends the asymptotic giant branch 
phase would require exceptionally fine-tuned migration within the 
stellar envelope \citep[see Section 5.3 of][]{cametal2019},
because the in-spiral time is at most $10^4$ orbits \citep{macetal2018}.

\subsection{Scattering the planet}

Assuming that the planet initially resides at a distance of at least 2-3 au, then it will need to be perturbed
by another planet of approximately equal or greater mass to eventually reach a distance within 0.07 au.
Many investigations have now shown that such inward scattering within evolved single-star multi-planet systems is feasible
\citep{debsig2002,veretal2013,voyetal2013,musetal2014,vergan2015,veretal2016,musetal2018,veretal2018b}.

{\rev A different potential scatterer could be a stellar companion, which has been shown to easily perturb major planets in evolved planetary systems close to one of the stars \citep{hampor2016,steetal2017,veretal2017b,steetal2018} during the post-main-sequence phases. However, WD~J0914+1914 does not have a known stellar binary companion. Further \cite{ganetal2019} ruled out both short-period companions and companions brighter than L5 dwarfs, the latter due to the absence of an infrared excess.}

Therefore, {\rev in the absence of a detected stellar companion} we conclude that WD~J0914+1914 contains or recently contained (within the last 13 Myr) 
at least one other major planet.
The location of these other planets depend on the details of the scattering and the dynamical instability.
As the above-cited studies demonstrated, the scattering represents a delayed effect 
triggered by dynamical instability which is itself instigated from stellar mass loss during the giant branch phases.
For WD~J0914+1914, the delay could not have lasted a long time: the respective durations of the red giant 
phase and asymptotic giant phases of this star
were, respectively, hundreds of Myr and several Myr (whereas the cooling age of WD~J0914+1914 is
in-between).

The scattering event needed to perturb WD~J0914+1914~b into an orbit with a pericentre which
is at most 0.07 au (and hence with an eccentricity of at least 0.93). 
{\rev Hence the planet may still reside on an eccentric orbit.}

An orbit which perhaps better fits the observations is one of a more circular nature.
Reducing the eccentricity temporarily could be achieved with a sustained scattering
scenario involving multiple planets and angular momentum transfer through 
the angular momentum deficit \citep{laskar1997,laskar2000,laspet2017} and 
secular chaos \citep{litwu2011,wulit2011,litwu2014}. 
However, in these scenarios, usually the innermost planet's eccentricity is only permanently
reduced only after tidal interactions with the star.

\subsection{Chaotic tidal evolution}

Tidal interactions between stars and planets can reduce the eccentricity of an orbit.
If, during this process, angular momentum
was conserved, then the quantity $a \left(1 - e^2\right) = q\left(1+e\right)$, where
$q$ is the orbital pericentre,
would remain constant. Hence, the periastron distance changes by a factor of $\left(1+e\right)$
as the eccentricity changes, and circularizing from $e=1$ to $e=0$ would double $q$. 
This scenario illustrates that the initial orbital pericentre of WD~J0914+1914~b
subsequent to scattering would have been located at about 0.035~au.

What is the timescale for this decrease in semimajor axis and increase in pericentre?
Constraints on the answer may be provided by dedicated tidal studies between white dwarfs and major planets,
investigations which are
only starting to emerge \citep{veretal2019,verwol2019,verful2019}.
Nevertheless, with the discovery of WD~J0914+1914~b, {\rev and potential for more discoveries
of the same type \citep{schetal2019}}, additional tidal studies may
be warranted.

For now, we utilize the results of \cite{verful2019}, which considered two
types of tidal interaction between white dwarfs and, specifically, gaseous planets.
The first (``chaotic tides'') is when a gaseous planet experiences pericentre encounters with
the white dwarf when on a very highly eccentric ($1-e \ll 0.1$) orbit.
The second (``non-chaotic tides'') occurs for lower eccentricities.
Like for WD~J0914+1914~b, major planets orbiting white dwarfs at small distances
will initially have eccentricities much higher than 0.9. Therefore, understanding and exploring
whether chaotic tides ``activates'' is the first task.

``Chaotic tides'' is shorthand for the stochastic orbital evolution which
occurs due to excitation of f-modes within the planet and the resulting exchange
of energy between those modes and the angular orbital momentum
\citep{mardling1995a,mardling1995b,ivapap2004,ivapap2007,viclai2018,wu2018,tesetal2019,vicetal2019}.
The effects of chaotic tides is drastic, sometimes allowing the semimajor axis to drop by over
90 per cent while changing the eccentricity by only a few hundredths in under 1 Myr
(see e.g. Fig. 1 of \citealt*{verful2019}). Crucially, chaotic tides also produce a negligible
change in the orbital pericentre.

\begin{figure*}
\includegraphics[width=17cm]{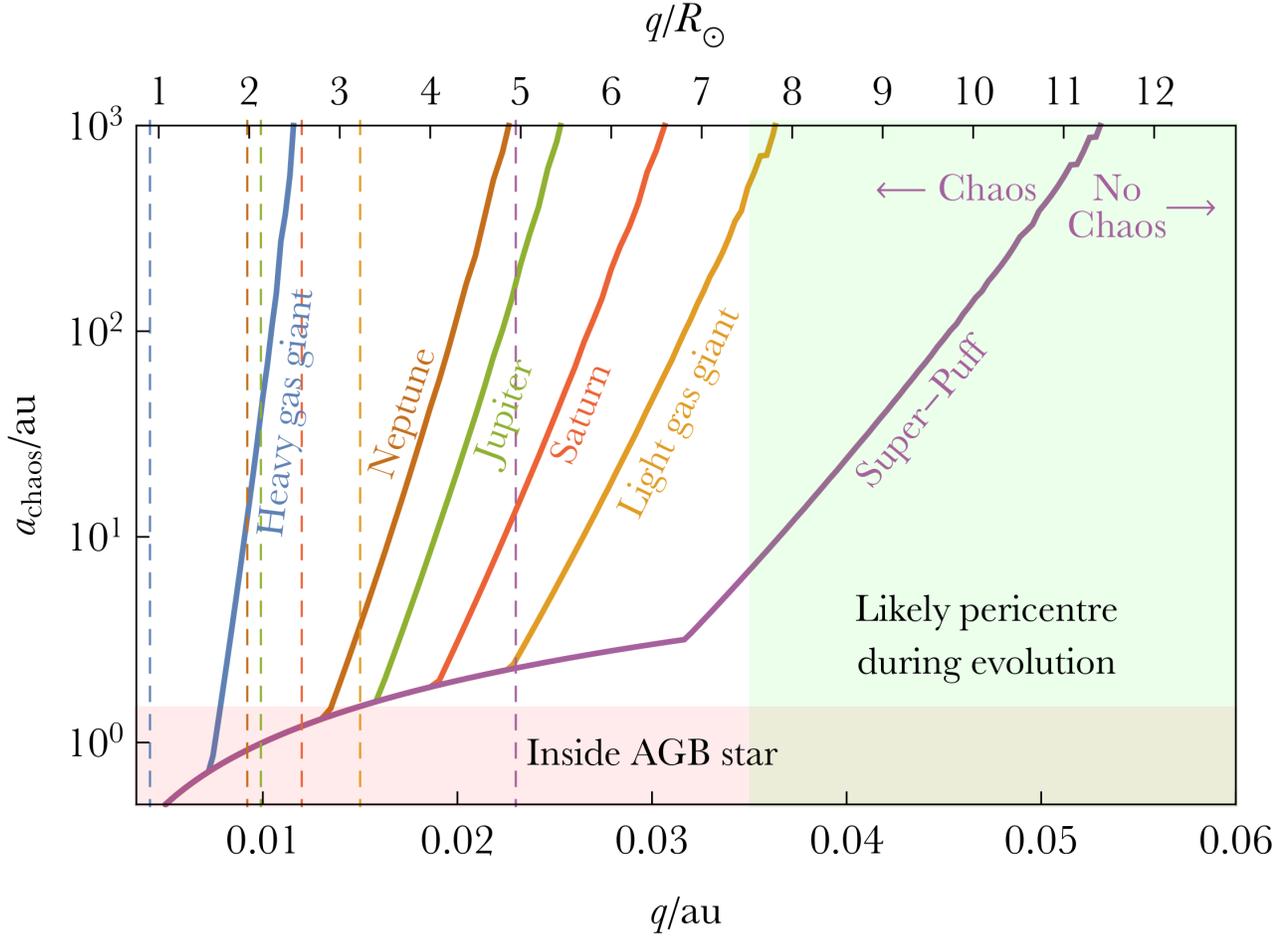}
\caption{
Demonstration that WD~J0914+1914~b was very unlikely to have experienced
chaotic tidal evolution unless the planet is or was a highly-inflated Super-Puff. The {\rev solid} curves 
represent the minimum initial semimajor axes
($y$-axis) for which different types of planets would have experienced chaotic tidal evolution
around the white dwarf for given orbital pericentres ($x$-axes). {\rev The vertical dashed lines are
representative white dwarf tidal disruption radii for each type of planet, ordered from left to right in the same way as the solid curves.} Because Super-Puffs are
particularly vunerable to self-disruption through chaotic tides, current observations may be
of a partially or fully disrupted ice giant.
}
\label{nochaos}
\end{figure*}

\begin{figure*}
\includegraphics[width=17cm]{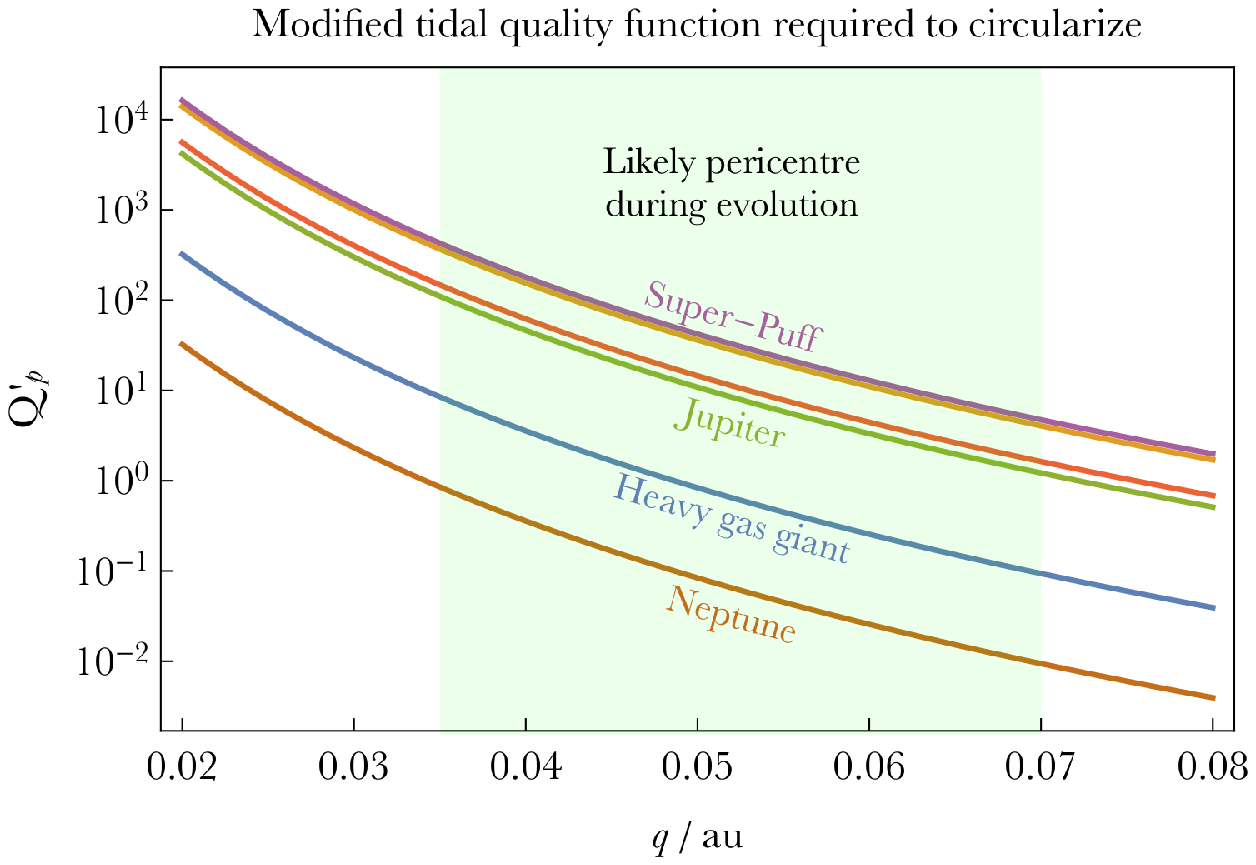}
\caption{
Modified tidal quality factors ($y$-axis) required to circularize the orbit of 
WD~J0914+1914~b depending on its physical parameters (various curves)
as a function of the orbital pericentre ($x$-axis). Potential values of the 
tidal quality factors are unrealistically small within the green
shaded region, which represents the likely range of $q$ values the planet has acquired
during white dwarf cooling if the planet's current pericentre resides at 0.07 au 
(as assumed in G\"{a}nsicke et al. 2019).
}
\label{Qtides}
\end{figure*}

This type of evolution would facilitate explanation of the dynamical origin
of WD~J0914+1914~b. However, we now argue that if the pericentre of
the planet was 0.035 au, then that value renders chaotic evolution impossible
unless the planet was a ``Super-Puff'' (highly inflated; nomenclature from \citealt*{leechi2016}).

A criterion for initiating chaotic evolution was given in 
Eq. 28 of \cite{vicetal2019}, and was re-expressed in Eq. 13
of \cite{verful2019}. We take the latter expression and
simplify it\footnote{For all of the detailed assumptions which enter into this criterion,
see \cite{vicetal2019}. One of these assumptions is a value for the overlap integral, which depends
on the structure of the planet. In particular, the value may be smaller for Super-Puffs -- which probably have low-mass envelopes -- than gas giants. We adopt the same value for the integral (0.56) as in \cite{vicetal2019} and \cite{verful2019}.} further to

\begin{equation}
116.7 \frac
{\sigma^2 K^2}
{\epsilon}
>
\left( 
\frac{M_{\rm p}}{M_{\star}}
\right)
\frac
{a^{7/2} \left(1 - e\right)^6\sqrt{G \left(M_{\star} + M_{\rm p} \right) }}
{R_{\rm p}^5}
,
\label{critchaos}
\end{equation}

\noindent{where} the variables on the right-hand-side of the equation
are standard ($a$ for the semimajor axis, $e$ for the eccentricity,
$M_{\star}$ and $M_{\rm p}$ for the mass of the star and
planet, and $R_{\rm p}$ for the radius of the planet). The variables
on the left-hand side are the f-mode frequency

\begin{equation}
\epsilon = 1.22 \sqrt{\frac{GM_{\rm p}}{R_{\rm p}^3}}
,
\end{equation}

\noindent{}the {\rev spin frequency of the planet, which is assumed to rotate pseudosynchronously according to}

\begin{equation}
\Omega = \frac
{1 + \frac{15}{2}e^2 + \frac{45}{8}e^4 + \frac{5}{16}e^6}
{\left(1 - e^2\right)^{3/2} \left(1 + 3e^2 + \frac{3}{8}e^4 \right)}
\sqrt{\frac{G\left(M_{\star} + M_{\rm p}\right)}{a^3}}
,
\end{equation}

\noindent{and}

\begin{equation}
\sigma = \epsilon + \Omega
,
\end{equation}

\noindent{}where

\[
K = 0.52 z^{3/2} \left[{\rm exp}\left(-\frac{2}{3}z \right)\right]
\left(1 - \frac{0.44}{\sqrt{z}} \right)
\]

\begin{equation}
\ \ \ \ 
\times
\left[
\frac{a\left(1 - e\right)}{R_{\rm p}}
\right]^{3/2}
\sqrt{\frac{M_{\rm p}}{M_{\star}}}
,
\end{equation}

\noindent{}with

\begin{equation}
z = \sqrt{2} \frac{\sigma}{\Omega}
.
\label{zequ}
\end{equation}

\cite{verful2019} found that whether or not chaotic evolution is activated
depends strongly on the orbital pericentre $q$, {\rev and can occur only when
$u \equiv q/r_{\rm Roche} \lesssim 2.0$, where $r_{\rm Roche}$ is the tidal disruption radius
of the white dwarf.} Here, we numerically solved 
the set of equations (\ref{critchaos})-(\ref{zequ}) by computing
the critical values of $a$, denoted by $a_{\rm chaos}$, which lead to an equality in equation (\ref{critchaos}) for 
given values of $q$. The result is illustrated in Fig. \ref{nochaos}.

On the figure, we computed critical curves for six different types of planets, helping to bound 
the entire plausible range for the initiation of chaotic tides. {\rev We assumed that the white
dwarf tidal disruption radius took on the same form as in \cite{verful2019}:

\begin{equation}
r_{\rm Roche} = 1.619 R_{\odot} \left( \frac{\rho_{\rm p}}{3 \ {\rm g} \ {\rm cm}^{-3}} \right)^{-1/3}
\end{equation}

\noindent{}where $\rho_{\rm p}$ is the density of the planet. A more accurate expression for the tidal disruption 
radius would depend on the planet's physical characteristics, including its spin, and hence would change along with the spin after each pericentre passage. The six types of planets, presented
along with their $u$ and $r_{\rm Roche}$ values (assuming $q = 0.035$ au), are

\begin{enumerate}

\item An exo-Jupiter ($u = 3.54, r_{\rm Roche} = 0.0099$ au),

\item An exo-Saturn ($u = 2.84, r_{\rm Roche} = 0.012$ au),

\item An exo-Neptune ($u = 3.80, r_{\rm Roche} = 0.0092$ au),

\item  A ``heavy gas giant'' ($M_{\rm p} = 13M_{\rm Jupiter}$ and $R_{\rm p} = R_{\rm Jupiter}$, giving $u = 8.33, r_{\rm Roche} = 0.0042$ au),

\item A ``light gas giant'' ($M_{\rm p} = 0.3M_{\rm Jupiter}$ and $R_{\rm p} = R_{\rm Jupiter}$, giving $u = 2.37, r_{\rm Roche} = 0.015$ au),

\item A ``Super-Puff'' ($M_{\rm p} = 4M_\oplus$ and $R_{\rm p} = 6R_\oplus$, giving $u=1.51, r_{\rm Roche} = 0.023$ au)

\end{enumerate}

}
Chaotic evolution is activated only for semimajor axes above the curves. Also plotted {\rev
are representative white dwarf tidal disruption radii for the different types of planets as dashed vertical lines}, as well as an approximate
value for the asymptotic giant branch radius for WD~J0914+1914 (recall that tidal
engulfment would occur at {\rev different values}, again depending on the planet characteristics).

Fig. \ref{nochaos} clearly demonstrates that a pericentre of 0.035~au 
is too high to have initiated
chaotic evolution in the past unless the planet was a highly-inflated ``Super-Puff''. {\rev
This result also conforms with the findings from \cite{verful2019} that chaotic tides
is not activated when $u \gtrsim 2.0$.}
For the Super-Puff case, the semimajor axis of the planet needed to be at least about 10 au to 
initiate chaotic tides.

Such an inflated planet would have likely experienced multiple thermalization
events during the chaotic evolution. As shown in \cite{verful2019}, exo-Neptune
analouges would be already susceptible to self-disruption through these thermalization events.
For a Super-Puff, the self-disruption would occur sooner. However, the disruption process has
not yet been analyzed in detail, and current observations might plausibly reveal 
an icy or rocky core of a disrupted planet.

\subsection{Non-chaotic tidal evolution}

If chaotic tidal evolution did not occur, then the planet's orbital and physical evolution was
qualitatively different.  In order to explore non-chaotic
tidal evolution, we adopt the simple and standard equilibrium weak friction tidal approximation from 
\cite{hut1981}.

In this approximation, the planet's semimajor axis and eccentricity evolve according to a set of differential
equations which are a function of the modified quality factors of the planet and star.
\cite{verful2019} found that to a good approximation for white dwarf planetary systems,
dissipation in the white dwarf can be neglected and the timescale to circularize the orbit, $\tau_{\rm circ}$, 
can be empirically estimated through\footnote{In order to obtain the formula, they assumed a white dwarf mass of $0.60M_{\odot}$,
which is sufficiently similar to the mass of WD~J0914+1914 ($0.56M_{\odot}$) for us to use here.}

\[
\tau_{\rm circ} \approx 37.4 \ {\rm Myr} \left( \frac{q}{r_{\rm Roche}} \right)^{13/2} 
                                             \left( \frac{Q_{\rm p}'}{10^6}  \right)
\]

\begin{equation}
\ \ \ \ \ \ \times                                             \left( \frac{M_{\rm p}}{M_{\rm Jupiter}}  \right)^{-2/3}
                                             \left( \frac{\rho_{\rm p}}{1 \ {\rm g/cm}^3}  \right)^{-1/2},
\label{nonchaos}
\end{equation}

\noindent{}where $Q_{\rm p}'$ is the modified planetary quality factor and $\rho_{\rm p}$ is the planet's
density. 

Because equation (\ref{nonchaos}) showcases a particularly strong dependence on $q$,
we plot values of  $Q_{\rm p}'$ which yield $\tau_{\rm circ}= 13.3$ Myr (the cooling age of WD~J0914+1914)
as a function of $q$ in Fig. \ref{Qtides}. The region with the green shaded background represents
the likely range of $q$ values that the orbit of WD~J0914+1914~b has acquired throughout white dwarf cooling. 
Note that regardless of the planet's $q$ value in this region, the required
values of $Q_{\rm p}'$ are typically orders of magnitude lower than the typically considered range
of $10^3-10^7$ \citep{wu2005,matetal2010,ogilvie2014}. The planet's current $q$ would have to be less than half of the estimated value of 0.07 au in order for weakly dissipative tides to 
represent a more plausible circularization mechanism.

\section{Discussion}

Our results suggest that a partially or fully intact WD~J0914+1914~b cannot reside on a near-circular 
orbit of about 0.07 au unless the planet is (or was) a Super-Puff (highly inflated) which has undergone
chaotic tidal evolution. Nevertheless, our understanding of tidal interactions still leaves
much room for improvement. \cite{efrmak2013} illustrated that popular tidal
models are unphysical in many situations, and definitive observational confirmation of tidally-induced
orbital decay in extrasolar systems
is only now becoming a reality {\rev \citep{macetal2013,hoyetal2016,patetal2017,wiletal2017,baigoo2019,yeeetal2019}}.

If the planet is not a Super-Puff and does have a pericentre distance of 0.07 au, then it 
would need to {\rev reside} on an eccentric orbit; observations
cannot yet support or refute this possibility. If this orbital eccentricity is sufficiently high,
then it would interact repeatedly with the gaseous disc which is thought to extend out to 0.04-0.05~au.
The consequences of this interaction depend on the geometric details of the disc and the relative inclination of the planet's
orbit to the disc plane. The picture is complicated further by the evaporation of the planet's atmosphere,
which would be a strong function of distance and be most significant close to pericentre \citep{schetal2019}. 
The potential interaction between
the planet and gas disc strongly motivates follow-up observations, particularly because, as estimated earlier,
the duration of pericentre passages is on the order of just days.

Instead, a partially or fully disrupted planet motivates further modelling of the energy redistribution and loss
within the f-modes of a planet. The destination of the energy after each 
thermalization event remains unknown and is dictated by the subtleties of the 
non-linear breaking process, which would again require detailed modelling. 
If the energy is assumed to be retained within the planet,
then eventually the planet could self-disrupt. Even before this time, a 
fragile atmosphere might leak out due to only a few thermalization events.

\cite{verful2019} illustrated that the timescale for potential self-disruption
for ice giants with the same mass and radius as Neptune is less than about 10~Myr as long as the orbital pericentre is within
about $1.5r_{\rm Roche} \approx 0.014$~au $\approx 2R_{\odot}$.
As shown above, it is theoretically difficult to circularize a planet to a 
semi-major axis as large as $\approx 0.07$~au (as inferred by \citealt*{ganetal2019}) 
within the cooling age (13 Myr) of this host star. One possibility is that chaotic tidal migration of 
a Super-Puff planet began at smaller periastron distance, but the subsequent heating 
disrupted the planet (or stripped its atmosphere) to produce the debris disk around WD J0914+1914.
Another possibility is that the core of such a partially disrupted planet could remain intact and 
circularize to a small semi-major axis upon interaction with the disruption debris. 

A third possibility is that the current orbit and physical state of WD~J0914+1914~b is the result of a giant impact between two major planets; the white dwarf, being so young, could have easily harboured a violent dynamical environment.
This impact --- which may have occurred around 0.07~au --- could have reduced the eccentricity of the planet's orbit, 
as well as produced a stream of debris. The high luminosity of this newly-formed white dwarf could have then generated
an inward drag of the resulting debris \citep{stoetal2015,veretal2015b}. Over time, in combination with evaporation from the planet's
atmosphere, the interaction between the planet and the debris may have cleared out the region in-between the disc and planet (between about $0.045$ and $0.070$ au).

A final possibility is that the semi-major axis of WD J0914+1914~b is smaller than the 0.07 au estimated by \cite{ganetal2019}. If the planet's mass is less than about $20 M_{\oplus}$, then it may not be massive enough to open a gap nor truncate the accretion disc. Hence, the planet could orbit {\it within} the debris disc or at its outer edge with a semi-major axis of $\approx 0.04$ au. In this scenario, the initial periastron distance of the planet may have been closer to 0.02 au, increasing the plausibility of both the chaotic and equilibrium tidal circularization phases. A prediction of this model could be tested through periodic distortions in the shape of the emission lines of the disc, similar to those observed in the debris disc around SDSS J1228+1040 \citep{manetal2019}. 

{\rev Despite the fine details of the planet's current orbit, the planet's classification as a Super-Puff would be consistent with the gravitational scattering hypothesis. The planet could have been born at a sufficiently large semimajor axis under a wide range of nebular conditions \citep{leeetal2018} to retain its puffy atmosphere throughout stellar evolution \citep{wandai2019}. Subsequently, the planet may have migrated inward \citep{leechi2016} under the high-eccentricity regime to arrive at its current location.}

 \section{Summary}

The robust signatures of the first major planet found orbiting a single white dwarf \citep{ganetal2019}
reveal a planet which is better constrained chemically than dynamically. Observations strongly suggest that
the planet is an ice giant (or the remnants of one), but its mass, radius and orbit are unknown, except for an 
inferred but still uncertain current distance of about 0.07 au. 
Contrastingly, the cooling age of the planet's host star is well-constrained ($13.3 \pm 0.5$ Myr) 
and in fact is much better constrained than the age of almost any other known major planet host star.

The juxtaposition of such robust and poor constraints requires an unorthodox analysis
to identify the dynamical history of WD~J0914+1914~b.
Here we considered scenarios which can explain the observations by assuming 
the inferred current distance of 0.07 au. We claim
that the planet must have resided at a distance of at 
least a few au at the onset of the white dwarf phase, and subsequently been scattered 
towards the white dwarf with a pericentre which is about 0.035~au. This scattering event 
requires the current or former presence of at least
one other major planet in the system of comparable or greater mass.

We also suggest that tidal circularization at 0.07 au could 
not have occurred unless the planet is a highly-inflated ``Super-Puff''.
In the distance range $0.035-0.070$ au, weak equilibrium
tides require unrealistically low values for the planetary quality factor to circularize an orbit
within the white dwarf's cooling age of 13 Myr. Instead, chaotic f-mode tidal
evolution would be required, which could shrink the semimajor axis orders of magnitude more quickly.

Further, the inflated nature of this planet implies,
from \cite{verful2019}, that the planet would have experienced thermalization events during this chaotic tidal evolution. These events could have partially or fully disrupted the planet. Alternatively, a denser or 
larger ice giant may be intact and passing through 0.07 au on its way to a much
lower pericentre. Finally, a remaining possibility is that an intact planet orbits {\it within} the debris disk at 
$\approx 0.04$ au, allowing for tidal circularization to deliver the planet to its current location within the short cooling age of the white dwarf.

These varied and violent possibilities provide
strong motivation for the acquisition of future observational data of WD~J0914+1914~b.

\section*{Acknowledgements}

{\rev We thank the referee for helpful comments which have improved the manuscript, and also 
thank Uri Malamud and Alexander Stephan for useful discussions.}
DV gratefully acknowledges the support of the STFC via an Ernest Rutherford Fellowship (grant ST/P003850/1). JF acknowledges support from an Innovator Grant from The Rose Hills Foundation and the Sloan Foundation through grant FG-2018-10515.

\label{lastpage}
\end{document}